\documentclass{article}
\usepackage{hyperref}
\usepackage[left=2.5cm,right=2.5cm,top=2.5cm,bottom=2.5cm]{geometry}
\usepackage{graphicx}
\usepackage{amsmath}
\usepackage{amssymb}
\usepackage{epstopdf}

\usepackage{color}
\usepackage{xcolor}
\newcommand{\fref}[1]{Fig.~\ref{#1}}
\newcommand{\Fref}[1]{Figure~\ref{#1}}

\newcommand{\sref}[1]{Sec.~\ref{#1}}

\newcommand{\Z}{\mathbb{Z}}

\begin{document}

\title{Semiclassical bifurcations and quantum trajectories: a case study of the open Bose-Hubbard dimer}

\author{Andrus Giraldo$^{1,2,}$, Stuart J. Masson$^3$, Neil G.R. Broderick$^{1,4}$  and \\ Bernd Krauskopf$^{1,2}$ \\[0.5cm] $^1$ Dodd-Walls Centre for Photonic and Quantum Technologies, New Zealand \\
$^2$ Department of Mathematics, University of Auckland, Auckland 1142, New Zealand \\
$^3$ Department of Physics, Columbia University, New York, NY 10027, USA \\
$^4$ Department of Physics, University of Auckland, Auckland 1142, New Zealand}

\date{September 2021}

\maketitle

\begin{abstract}
We consider the open two-site Bose-Hubbard dimer, a well-known quantum mechanical model that has been realised recently for photons in two coupled photonic crystal nanocavities. The system is described by a Lindblad master equation which, for large numbers of photons, gives rise to a limiting semiclassical model in the form of a four-dimensional vector field. From the situation where both sites trap the same amount of photons under symmetric pumping,  one encounters a transition that involves symmetry breaking, the creation of periodic oscillations and multistability as the pump strength is increased. We show that the associated one-parameter bifurcation diagram of the semiclassical model captures the essence of statistical properties of computed quantum trajectories as the pump strength is increased. Even for small numbers of photons, the fingerprint of the semiclassical bifurcations can be recognised reliably in observables of quantum trajectories.
\end{abstract}

\section{Introduction} 
\label{sec:Intro}

Phase transitions describe a fundamental change in the behavior of a system as its parameters are changed. Understanding why these systems exhibit different observable features in different physical scenarios and how they transition between them is a fundamental problem in many branches of physics. In quantum systems, the interest is in phase transitions, which occur near zero temperature and are driven by quantum fluctuations~\cite{Greiner:2002wt,Heyl_2018}. We are concerned here with an open quantum system, where particles of an ensemble may be lost to the environment. Open quantum systems constitute a realistic scenario from an experimental perspective, where lost particles can be measured to monitor the system as it evolves. The underlying non-unitary evolution can produce dissipation and decoherence not exhibited by their closed counterparts.

For both open and closed quantum systems, quantum fluctuations can become negligible in the thermodynamic limit of large particle number (e.g. atoms and/or photons). Their time evolution can be described by a so-called \textit{semiclassical model} or mean-field approximation \cite{OzoriodeAlmeidaA.2011}, which takes the form of a closed set of ordinary differential equations (ODEs) for associated averaged quantities. Such an approach has long been common in quantum optics to study systems with large numbers of photons, such as lasers, and it is the starting point of our analysis here. The attractors of the semiclassical ODE, that is, its stable solutions, therefore, give insight into the observable behaviour of the underlying quantum system. In other words, phase transitions exhibited by the quantum system can be identified as changes in the stability of solutions of the limiting semiclassical model, which are examples of \textit{bifurcations}. Thus, bifurcation analysis of the semiclassical ODE --- with a combination of analytical and advanced numerical tools --- allows one to systematically map out the attractors and phase transitions in the thermodynamical limit; see for example, \cite{guckenheimer1983nonlinear,Kuz1,redbook} as entry points to bifurcation theory and associated numerical methods. It has long been common in quantum optics to study systems with large numbers of photons, such as lasers, by means of their semiclassical descriptions. 

We are interested here in the relationship between dynamics and bifurcations of the semiclassical ODE model and the observable behaviour of the quantum system when the number of particles of the considered system is relatively small. In this case, the system is far from the thermodynamic limit, and quantum fluctuations may be significant and cannot simply be neglected. Quantum simulations of the underlying Hamiltonian, which are only feasible computationally for small numbers of particles, can be used to investigate the predictive power of semiclassical models for novel types of quantum systems that operate with very few atoms and/or photons.  It has been shown by means of quantum simulations that fingerprints of semiclassical predictions can still be found in probabilistic features of quantum systems for surprisingly low numbers of particles~\cite{Carmichael15,PhysRevA.98.023804,Vukics19,PhysRevResearch.Kevin2021}.  As a recent example, the semiclassical limit of the unbalanced Dicke model  \cite{PhysRevA.Kevin2021} predicts parameter regimes of superradiant switching, quantum hysteresis, and oscillations which have been observed in its quantum optical description \cite{PhysRevResearch.Kevin2021}.

This paper probes the connections between the semiclassical and quantum regime for the specific example of the open two-site Bose-Hubbard dimer \cite{someBHrefs1,someBHrefs2,Kordas2015}. This well-known quantum mechanical model describes the dynamics of bosonic particles in a lattice, where the behaviour is determined by the interplay of the hopping rate of particles between lattice sites and their on-site interactions. Different experimental realisations of this system have been achieved in the form of semiconductor microcavities~\cite{lagoudakis_coherent_2010,abbarchi_macroscopic_2013,rodriguez_interaction-induced_2016} and superconducting circuits~\cite{raftery_observation_2014,eichler_quantum-limited_2014}. Of particular interest to us is an optical realisation in the form of two lossy coupled photonic crystal resonators~\cite{Hamel_2015,garbin2021spontaneous}. In this setup, good agreement between experimental measurements and the semiclassical description has been obtained in a regime up to moderate strength of the optical drive signal, in particular, concerning the observation of spontaneous symmetry breaking~\cite{GBKYA_2020}. Recent theoretical results over a much wider range of parameters have uncovered a rich variety of behaviour of the  semiclassical open Bose-Hubbard dimer, including chaotic behaviour with non-switching, regular switching and chaotic switching between the two sites; see \cite{GBK_2021} for more details. We focus here specifically on the comparison between the quantum and semiclassical descriptions of the two-site Bose-Hubbard model. Such comparisons have been performed before under the excitation of the anti-bonding mode~\cite{Casteels2017}, and under excitation of the bonding mode for the case of positive intermode coupling~\cite{BinMahmud,Lledo2019}. Here we consider the situation where the bonding mode is only excited while the intermode coupling of the cavities is negative. In this way, we extend earlier comparisons to the specific parameter constellations that most resembles the experimental setup in \cite{garbin2021spontaneous} and, to our knowledge, have not been studied before.

Our approach is as follows: we choose parameter values from \cite{GBKYA_2020} to obtain a bifurcation diagram of the semiclassical ODE model with different qualitative behaviour, specifically, symmetry breaking, periodic motion and multi-stability of asymmetric states. We then obtain temporal traces of the open Bose--Hubbard dimer by means of quantum trajectory simulations \cite{CarmichaelHoward2007Smiq}. We then compare the two. By performing simulations of the quantum system for an increasing number of photons, we are able to observe the emergence of semiclassical attractors in a fully quantum realisation even with low photon number. Furthermore, we consider signatures for the existence of anti-bunching \cite{PhysRevLett.108.183601} and entanglement \cite{PhysRevLett.2004_Howard,PhysRevLett.88.197901} in the quantum system as different bifurcations of the semiclassical approximation are encountered. Overall, our results showcase the role a semiclassical bifurcation analysis of a quantum system may play in providing a roadmap of interesting fundamental behaviour of interest, both from a theoretical and an experimental perspective. Specifically for the open two-site Bose--Hubbard dimer, this demonstrates that it might be feasible to investigate the quantum footprint of even quite complex dynamical behaviour, such as different types of chaotic switching~\cite{GBK_2021}.

The computations of equilibria and local bifurcations presented here are implemented in and performed with the pseudo-arclength continuation package \textsc{Auto07p} \cite{Doe1,Doe2} and its extension \textsc{HomCont} \cite{san2}.  The quantum trajectory simulations were carried out in the software package \textsc{QuTiP} \cite{JohanssonJ.Qutip2012,JohanssonJ.Qutip2013}. Visualisation and post-processing of the data are performed with \textsc{Matlab}\textsuperscript{\textregistered}.

\section{Quantum Hamiltonian and semiclassical ODE model}
\label{sec:BHmodel}

The Hamiltonian of the Bose--Hubbard model for two cavities takes the form \cite{BinMahmud} 
\begin{equation} \label{eq:Hamiltonian}
\begin{aligned}
\hat{H} = -J &(\hat{a}_1^{\dagger}\hat{a}_2 +
\hat{a}_2^{\dagger}\hat{a}_1 ) + \sum_{j=1,2}\left( \omega_c
  \hat{a}_j^{\dagger}\hat{a}_j + U
  \hat{a}_j^{\dagger}\hat{a}_j^{\dagger}\hat{a}_j \hat{a}_j  \right)
  \\
  &+ \sum_{j=1,2}\left( F e^{-i\omega_p t}\hat{a}_j + F^* e^{i\omega_p t}\hat{a}^{\dagger}_j \right),
\end{aligned}
\end{equation}
where $\hat{a}_j^{\dagger}$ is the creation operator at cavity $j$. The intermode coupling is represented by $J$,  the frequency inside the cavities by $\omega_c$ and the on-site energy by $U$. The last term of \eqref{eq:Hamiltonian} represents a driving field with frequency $\omega_p$ and amplitude $F$, which we consider equal for both cavities. In what follows, we work in a co-moving frame with respect of $\omega_p$; in this way, the time dependency of the Hamiltonian~\eqref{eq:Hamiltonian} is dropped. Since we are interested in two lossy mutually coupled cavities, the Lindblad master equation, which couples a thermal bath with the dynamics induced by \eqref{eq:Hamiltonian}, takes the form
\begin{equation} \label{eq:LindForm}
i\frac{d\hat{\rho}}{dt} = [ \hat{H}, \hat{\rho}] + i
\frac{\gamma}{2}\sum_{j=1,2}[2 \hat{a}_j \hat{\rho}
\hat{a}_j^{\dagger}-\hat{a}_j^{\dagger}\hat{a}_j \hat{\rho}-
\hat{\rho} \hat{a}_j^{\dagger}\hat{a}_j],
\end{equation}
where $\hat{\rho}$ is the density matrix and $\gamma$ is the loss rate. 

System~\eqref{eq:LindForm} can be evolved numerically by computing quantum trajectories~\cite{CarmichaelHoward2007Smiq}, where the wavefunction is evolved with a Monte Carlo algorithm. Such a computation is performed for a non-Hermitian effective Hamiltonian 
\begin{equation} \label{eq:Hamiltonian2}
\begin{aligned}
\hat{H} = -J &(\hat{a}_1^{\dagger}\hat{a}_2 +
\hat{a}_2^{\dagger}\hat{a}_1 ) + \sum_{j=1,2}\left( - \Delta \hat{a}_j^{\dagger}\hat{a}_j + U  \hat{a}_j^{\dagger}\hat{a}_j^{\dagger}\hat{a}_j \hat{a}_j  \right)
  \\
  &+ \sum_{j=1,2}\left( F \hat{a}_j + F^* \hat{a}^{\dagger}_j \right) - \frac{i\gamma}{2} \left( \hat{a}_1^\dagger \hat{a}_1+ \hat{a}_2^\dagger \hat{a}_2 \right).
\end{aligned}
\end{equation}
where the non-Hermitian terms account for null-measurement back-action; moreover, Eq.~\eqref{eq:Hamiltonian2} is expressed in a co-moving frame with respect to the frequency $\omega_p$ of the driving field,  and $\Delta=\omega_c-\omega_p$ is the detuning. The coherent evolution is interrupted by `jumps', which correspond to photon emission from the cavities implemented by the action of $\hat{a}_{1}$ or $\hat{a}_2$ on the wavefunction. The timing of these jumps is stochastic, and each trajectory corresponds to a particular \textit{unravelling} of the master equation given by system~\eqref{eq:LindForm}. This approach is quite powerful as one can obtain temporal information regarding observables, which mimics the situation of an observer taking measurements of the outputs of the cavities; see \cite{CarmichaelHoward2007Smiq} for more details.  We consider here the observables $\langle a^\dagger_1 a_1  \rangle$ and $\langle a^\dagger_2 a_2  \rangle$, the photon number expectation values in each of the two cavities.

In the co-moving frame with respect the driving frequency $\omega_p$,  the mean-field approximation
\begin{equation} \label{eq:Coupled}
\begin{aligned}
i\dfrac{d \alpha_1}{d \tau} & =  \left( -\Delta - i \dfrac{\gamma}{2}  +2U|\alpha_1|^2 \right) \alpha_1 -J \alpha_2  + F, \\
i\dfrac{d \alpha_2}{d \tau} & =  \left( -\Delta - i \dfrac{\gamma}{2}  +2U|\alpha_2|^2 \right) \alpha_2 - J \alpha_1  + F,  
\end{aligned}
\end{equation}
is derived from \eqref{eq:LindForm}  by considering the expectations (mean-values) $\alpha_1 = \langle \hat{a}_1 \rangle$ and $\alpha_2 = \langle \hat{a}_2 \rangle$ under the assumption that their product factorises, meaning that, for example, $\langle \hat{a}_1\hat{a}_2 \rangle =\langle \hat{a}_1\rangle \langle \hat{a}_2 \rangle$; see \cite{Kordas2015} for details. 
It is mathematically convenient to rescale time and the variables of \eqref{eq:Coupled} by introducing
$$\tau = 2t/\gamma  \quad \mbox{and} \quad
(A,B) =(-2i \overline{\alpha_1}\sqrt{|U|/\gamma},-2i \overline{\alpha_2}\sqrt{|U|/\gamma})$$
to obtain the system
\begin{equation} 
\label{eq:CoupledAB}
\begin{aligned}
\dfrac{d A}{d t} & = -A + i\left(\delta + \xi(U) |A|^2 \right) A +i \kappa B  + f, \\
\dfrac{d B}{d t} & =-B + i\left(\delta   + \xi(U) |B|^2 \right) B+ i \kappa A +f.  
\end{aligned}
\end{equation}
for the rescaled complex field electric amplitudes $A$ and $B$, where 
\begin{equation} 
\label{eq:parscalings}
\xi(U)=\text{sign}(U), \quad  \delta = -\frac{2\Delta}{\gamma},  \kappa=-\frac{2J}{\gamma} \quad \mbox{and} \quad f=4F\frac{\sqrt{|U|}}{\gamma^{3/2}}.
\end{equation}
System~\eqref{eq:CoupledAB} is the semiclassical ODE model of the open Bode-Hubbard dimer, and we will investigate how its bifurcations manifest themselves in the evolution as described by the Lindblad master equation~\eqref{eq:LindForm} for small photon numbers in the two cavities. Note that we are considering here the symmetric case where both cavities are pumped with the same intensity $F$; hence,  system~\eqref{eq:LindForm} is invariant under the exchange of the two cavities given by $(\hat{a}_1, \hat{a}_2) \rightarrow (\hat{a}_2, \hat{a}_1) $. This reflectional symmetry manifests itself as $\Z_2$-equivariance \cite{Golubitsky1985} of the semiclassical system~\eqref{eq:CoupledAB} with respect to the map $(A,B) \to (B,A)$.

Mathematically, system \eqref{eq:CoupledAB} is a system of four real-valued ODEs, which can be written out either in terms of real and imaginary parts or amplitudes and phases of both $A$ and $B$; hence, it defines a vector field with a four-dimensional phase space. The sign $\xi(U)=\text{sign}(U)$ appears in \eqref{eq:CoupledAB} to allow for convenient comparisons with different instances of the Bose-Hubbard dimer in the literature; however, the transformation
$$(A,B,U,\delta,\kappa) \mapsto (\overline{A},\overline{B},-U,-\delta,-\kappa)$$
implies that all results for $\xi(U)=1$ directly translate to those for $\xi(U)=-1$. From now on, we set $\xi(U)=\text{sign}(U)=1$ and consider positive $\kappa$ as the experimentally relevant case \cite{Hamel_2015,garbin2021spontaneous,Haddadi_2014} whose bifurcations have been studied in considerable detail in \cite{GBKYA_2020,GBK_2021}. More specifically, we also set $\delta=-4.5$ and consider the one-parameter bifurcation diagram in the pump strength $f$, for which one finds different types of phase transitions, including symmetry breaking, multistability and the onset of periodicity.

We compare the dynamics of the semiclassical ODE model~\eqref{eq:CoupledAB} with quantum trajectory computations, requiring that we set $\left( J,\Delta,U,\gamma\right) = \left( -3.5,4.5,0.5,2.0\right)$ in system~\eqref{eq:LindForm} so that the two sets of parameter values agree. Regarding pump strength $F$, we note that any scaling of the form 
\begin{equation} 
\label{eq:scaleF}
(U,F) \rightarrow \left(U_\mu,F_\mu\right) = \left(U / \mu,\sqrt{\mu} F\right)
\end{equation} 
with $\mu>0$ does not change the value of  $f$ in system~\eqref{eq:CoupledAB} as given by \eqref{eq:parscalings}; hence, the rescaling \eqref{eq:scaleF} leaves the observed semiclassical behaviour unchanged. Classically, this can be understood as simultaneously changing both the strength of the nonlinearity and the optical
intensity. On the other hand, in the quantum regime, the scaling factor $\mu$ plays a very important role: it encodes the number of photons in the cavities. As such, it enters as a scaling when one compares intensities; namely $\langle a^\dagger_1 a_1  \rangle=\mu |A|^2$ and $\langle a^\dagger_2 a_2  \rangle=\mu |B|^2$, which follows from the scaling in~\eqref{eq:parscalings} to obtain the pump strength $f$ of the semiclassical model~\eqref{eq:CoupledAB}. Increasing $\mu$ increases the number of photons and, hence, reduces the importance of cavity field fluctuations. In other words, the scaling factor $\mu$ allows one to simulate the quantum system for different numbers of photons trapped in the cavities, while the nature of the limiting semiclassical dynamics remains the same. In particular, by considering different values of $\mu$, one can investigate the importance of quantum fluctuations. We remark that this approach has been used in \cite{Lledo2019} to investigate system~\eqref{eq:Hamiltonian} for the case $J>0$.

\section{Bifurcation diagram and tracked quantum trajectory simulation} 
\label{sec:Phase}

\begin{figure}[t!]
   \centering
   \includegraphics{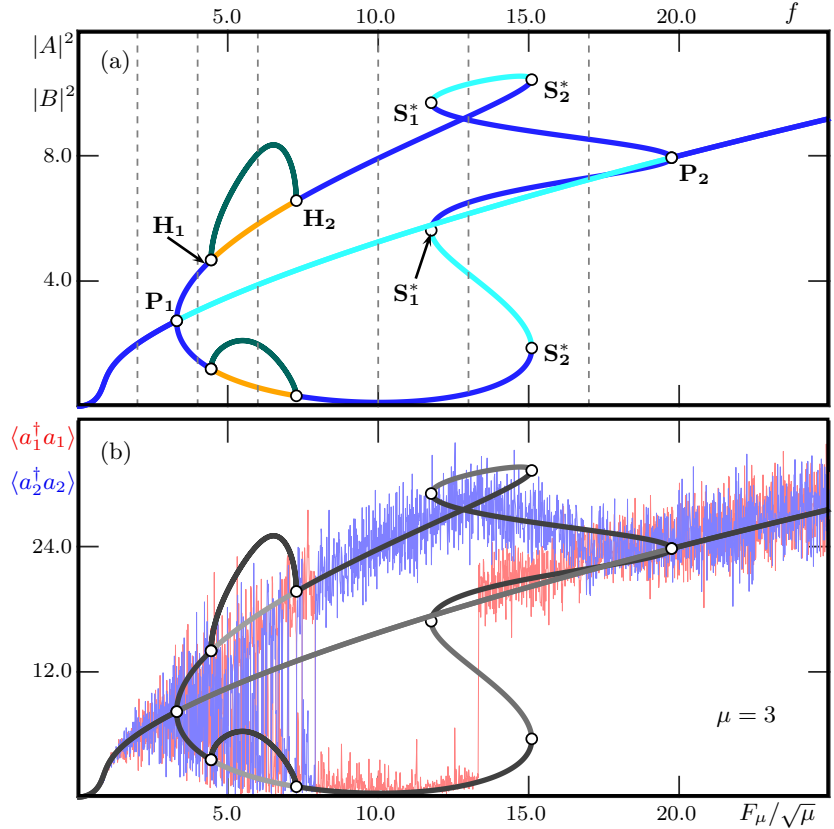}
    \caption{Bifurcation diagram of system~\eqref{eq:CoupledAB} in the pump strength $f$ compared with quantum trajectories of system~\eqref{eq:LindForm}. Panel~(a) shows branches of equilibria and periodic solutions represented by the intensities $|A|^2$ and $|B|^2$, namely: stable equilibria (blue), saddle equilibria with one unstable eigenvalue (cyan) and with two unstable eigenvalues (orange), and stable periodic solutions (dark green); these solutions bifurcate at points of pitchfork bifurcation $\mathbf{P}$, of saddle-node bifurcation of asymmetric equilibria $\mathbf{S_1^*}$ and $\mathbf{S_2^*}$, of saddle-node bifurcation of asymmetric equilibria  $\mathbf{S}^*$, and of Andronov--Hopf bifurcation $\mathbf{H}$. Panel~(b) shows quantum trajectory realisations of system~\eqref{eq:Hamiltonian2} as $F_\mu$ with $\mu=3.0$ is slowly ramped linearly as $F_{\mu}\approx 0.216\,t$; here the observables $\langle \hat{a}^{\dagger}_{1} \hat{a}_{1} \rangle$ (red curve) and $\langle \hat{a}^{\dagger}_{2} \hat{a}_{2} \rangle$ (blue curve) are superimposed on the branches of solutions from panel~(a) (dark and light grey curves). Throughout, $(\delta,\kappa)=(-4.5, 3.5)$ for system~\eqref{eq:CoupledAB} and $\left( J,\Delta,U,\gamma\right) = \left( -3.5,4.5,0.5,2.0\right)$ for system~\eqref{eq:Hamiltonian2}; the vertical grey-dashed lines indicate the $f$-values considered in \sref{sec:ComF}.}
\label{fig:BifDiag}
\end{figure}

We first compare in \fref{fig:BifDiag} the one-parameter bifurcation diagram in the pump current $f$ of system~\eqref{eq:CoupledAB} for $(\delta,\kappa)=(-4.5, 3.5)$ with quantum trajectories of system~\eqref{eq:LindForm} computed for slowly increasing pump current with corresponding parameter values. \Fref{fig:BifDiag}(a) shows the branches and bifurcations of the equilibrium and periodic solutions of system~\eqref{eq:CoupledAB} in terms of their intensities $|A|^2$ and $|B|^2$ over the relevant range of the pump strength $f$. Here, one finds stable equilibria along blue curves, while cyan and orange curves represent unstable equilibria with different numbers of positive eigenvalues. The green curves represent stable periodic solutions; specifically, these curves trace out the maxima (only) in $|A|^2$ and $|B|^2$ of these periodic solutions. Throughout, there exists a symmetric equilibrium with $|A|^2=|B|^2$ and $\arg{(A)}=\arg{(B)}$, and one finds a number of bifurcations that give rise to other branches of solutions. The $f$-range is divided in this way into intervals~\textbf{(i)}--\textbf{(vii)} with different qualitative behaviour as follows:
\begin{enumerate}
\item[\textbf{(i)}] 
$f=0$ to the pitchfork bifurcation $\mathbf{P_1}$: there exists only the symmetric equilibrium and it is stable.
\item[\textbf{(ii)}] 
$\mathbf{P_1}$ to the Andronov--Hopf bifurcation $\mathbf{H}$: the symmetric equilibrium is now unstable and a pair of stable asymmetric equilibria exist; these bifurcate at $\mathbf{P_1}$ with $|A|^2>|B|^2$ and $|A|^2<|B|^2$, respectively.
\item[\textbf{(iii)}] 
between the Andronov--Hopf bifurcations $\mathbf{H_1}$ and $\mathbf{H_2}$: 
the asymmetric equilibria are unstable, and there is now a pair of stable periodic orbits that emerge and disappear at $\mathbf{H_1}$ and $\mathbf{H_2}$ --- one near each equilibrium with $|A|^2>|B|^2$ and $|A|^2<|B|^2$, respectively---.
\item[\textbf{(iv)}]
$\mathbf{H_2}$ to the saddle-node bifurcation of asymmetric states $\mathbf{S^*_1}$: the asymmetric equilibria are again stable and the situation is as in interval~\textbf{(ii)}.
\item[\textbf{(v)}] 
between the saddle-node bifurcations of asymmetric states $\mathbf{S^*_1}$ and $\mathbf{S^*_2}$: two additional pairs of asymmetric equilibria exist, one stable and one unstable, so that there are now two pairs of stable asymmetric equilibria. The pair of unstable equilibria emerges or disappears with the respective pair of stable periodic orbits at the points $\mathbf{S^*_1}$ and $\mathbf{S^*_2}$.
\item[\textbf{(vi)}] 
$\mathbf{S^*_2}$ to the pitchfork bifurcation $\mathbf{P_2}$: there is again a single pair of stable asymmetric equilibria as well as the unstable symmetric equilibrium, as in intervals~\textbf{(ii)} and~\textbf{(iv)}.
\item[\textbf{(vii)}] 
beyond $\mathbf{P_2}$: the symmetric equilibrium is the only solution and stable again, as in interval~\textbf{(i)}.
\end{enumerate}

\Fref{fig:BifDiag}(b) shows the realisation of one quantum trajectory of system~\eqref{eq:LindForm} as $F_{\mu}$ with $\mu=3.0$ is increased linearly at a slow rate of approximately $0.216$ per unit of time $t$, where $\left( J,\Delta,U,\gamma\right) = \left( -3.5,4.5,0.5,2.0\right)$. The quantum trajectory is shown in terms of the two  observables $\langle a^\dagger_1 a_1  \rangle$ and $\langle a^\dagger_2 a_2 \rangle$, and superimposed is the bifurcation diagram from panel~(a) after the corresponding scaling by $\mu$ of the intensities $|A|^2$ and $|B|^2$; here stable branches are dark grey and unstable branches light grey. Notice in \fref{fig:BifDiag}(b) how both observables follow the branch of symmetric equilibria in interval~\textbf{(i)} and then split into an asymmetric situation when the semiclassical system exhibits the pitchfork bifurcation~$\mathbf{P_1}$. Near and beyond this transition, in intervals~\textbf{(i)} and~\textbf{(iii)}, there are increased fluctuations that include switching between which of $\langle a^\dagger_1 a_1  \rangle$ or $\langle a^\dagger_2 a_2 \rangle$ is the larger observable. Past the Andronov--Hopf bifurcation $\mathbf{H_2}$, in interval~\textbf{(iv)}, the quantum trajectory is clearly localised at one of the asymmetric stable equilibria; more precisely, the one with $\langle a^\dagger_2 a_2  \rangle>\langle a^\dagger_1 a_1  \rangle$. As $F_{\mu}$ increases, the quantum trajectory jumps to a different asymmetric equilibrium, namely in between the asymmetric saddle-node bifurcations $\mathbf{S^*_1}$ and $\mathbf{S^*_2}$ where two pairs of asymmetric equilibria exist, which is interval~\textbf{(v)}. As $F_{\mu}$ increases further, $\langle a^\dagger_1 a_1  \rangle$ and $\langle a^\dagger_2 a_2 \rangle$ remain localised near this new asymmetric equilibrium in interval~\textbf{(vi)}. The two observables  then come together near the pitchfork bifurcation $\mathbf{P_2}$ and then remain localised at the symmetric equilibrium that is stable again in parameter interval~\textbf{(vii)}.

Overall, \fref{fig:BifDiag} illustrates that the slowly ramped quantum trajectory effectively follows the stable branches of the one-parameter bifurcation diagram, meaning that it remains localised near one of the stable equilibrium solutions, but with a considerable level of fluctuations. Increased sensitivity and jumps induced by these fluctuations are observed near bifurcation points, in the $F_{\mu}$-range between $\mathbf{P_1}$ and $\mathbf{H_2}$ as well as that bounded by $\mathbf{S^*_1}$ and $\mathbf{S^*_2}$ with bistability between different asymmetric equilibria.

\section{Quantum trajectories at specific values of $f$} 
\label{sec:ComF}

We proceed by investigating how these observed properties of the quantum trajectory manifest themselves in the parameter intervals with different limiting semiclassical dynamics; specifically, at the values of $f = F_\mu/\sqrt{\mu}$ in intervals \textbf{(i)}--\textbf{(vi)}, as indicated by the dashed vertical lines in \fref{fig:BifDiag}(a). Here, we consider two values of the scaling factor $\mu$, that is, two different photon numbers, namely $\mu=1.0$ and $\mu=3.0$. More specifically, we present for each case:
\begin{itemize}
\item[(1)] 
temporal traces for a single realisation of a quantum trajectory of system~\eqref{eq:LindForm} represented by the observables $\langle a^\dagger_1a_1 \rangle$ and $\langle a^\dagger_2 a_2  \rangle$, shown over the corresponding range (that depends on $\mu$) with the respective equilibria of system~\eqref{eq:CoupledAB}.
\item[(2)]
the associated temporal traces of the ratio $$O=\dfrac{\langle a^\dagger_1 a^\dagger_2 a_1 a_2 \rangle}{\langle a^\dagger_1 a_1 \rangle \langle a^\dagger_2 a_2 \rangle},$$
which is a measure of the validity of the factorisation property of the quantum system used to derive the semiclassical ODE; note that $O=1$ means that the factorisation is exact.
\item[(3)] 
two-dimensional histogram in the $(\langle a^\dagger_1a_1 \rangle, \langle a^\dagger_2 a_2  \rangle)$-plane, where the bins are given by a $200\times200$ grid of the corresponding shown ranges and with the respective equilibria and periodic solutions of system~\eqref{eq:CoupledAB}. The histograms are constructed from three different realisations of quantum trajectories by sampling 50,000 equidistant points in the time interval $(0,10^4)$; moreover, the invariance of system~\eqref{eq:LindForm} under the permutation of the sites is used to double the number of points and to correct asymmetric bias introduced by the sampling.
\end{itemize}

\subsection{Comparison in intervals \textbf{(i)} to \textbf{(iii)}}
\label{sec:Int_i_to_iii}

\begin{figure}[t!]
   \centering
   \includegraphics{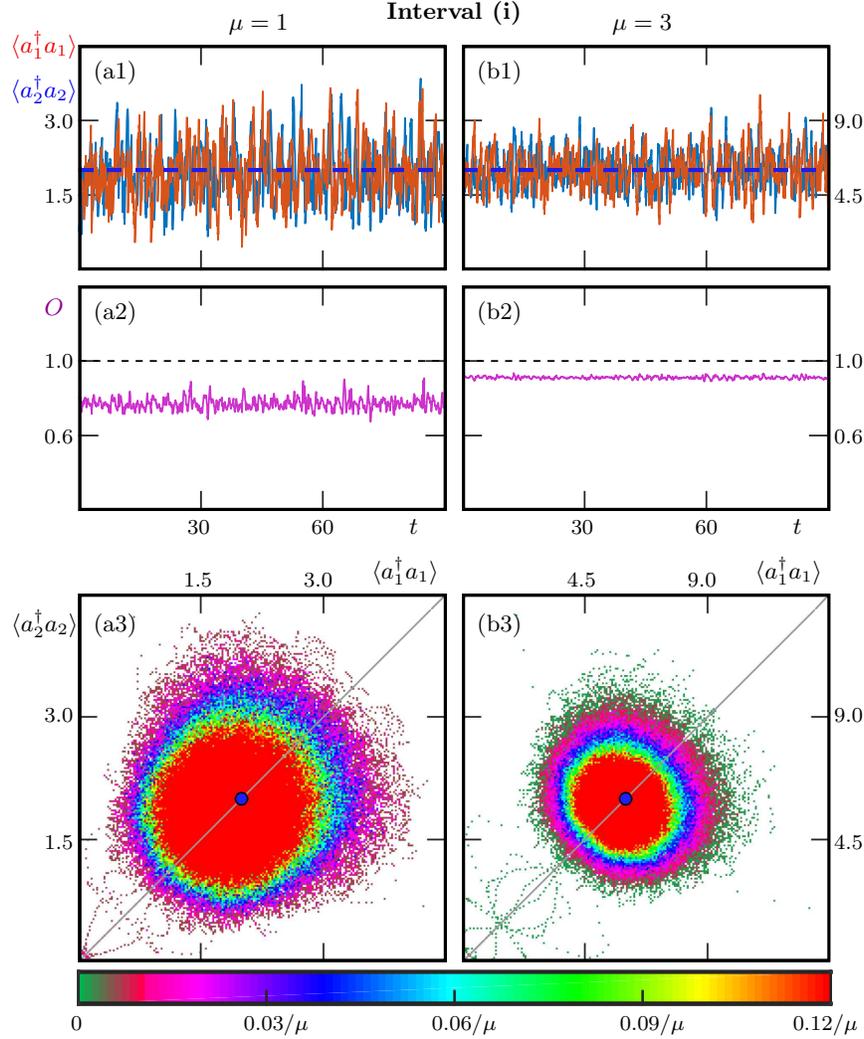}
    \caption{Quantum trajectories of system~\eqref{eq:LindForm} in interval~\textbf{(i)} for $F_\mu/\sqrt{\mu} = f = 2.0$ with $\mu=1.0$ (left column) and with $\mu=3.0$ (right column). Panels~(a1) and~(b1) show the temporal trace of the observables $\langle \hat{a}^{\dagger}_{1} \hat{a}_{1} \rangle$ (red curve) and $\langle \hat{a}^{\dagger}_{2} \hat{a}_{2} \rangle$ (blue curve), and panels~(a2) and~(b2) show the associated evolution of the ratio $O=\langle a^\dagger_1 a^\dagger_2 a_1 a_2 \rangle / \langle a^\dagger_1 a_1 \rangle \langle a^\dagger_2 a_2 \rangle$. Panels~(a3) and~(b3) show histograms for a $200\times200$ grid in the $(|A|^2, |B|^2)$-plane constructed from three different quantum trajectories; the data is symmetrised and the diagonal symmetry line is shown in grey. Also shown are the equilibria of system~\eqref{eq:CoupledAB}, as dashed lines in the temporeal traces and as dots in the histograms, where colour reflects their stability as in \fref{fig:BifDiag}.}
\label{fig:Quant_Int_i}
\end{figure}

In interval~\textbf{(i)} there exists only the single symmetric stable equilibrium of the semiclassical ODE~\eqref{eq:CoupledAB}, which is stable and attracts all initial conditions. As \fref{fig:Quant_Int_i} shows, the respective quantum trajectory stays close to this equilibrium but is subject to clear fluctuations. This localisation is illustrated in panels~(a1) and (b1) with the temporal traces of the two observables $\langle \hat{a}^{\dagger}_{1} \hat{a}_{1} \rangle$ and $\langle \hat{a}^{\dagger}_{2} \hat{a}_{2} \rangle$, which can be seen to fluctuate around the corresponding equilibrium intensity-value. Here the ranges have been chosen to agree with the scaling by $\mu$ so that a direct comparison of the observables can be made, including with their limiting semiclassical behaviour. Notice that the fluctuations are relatively larger for $\mu=1$ in panel~(a1) than those for $\mu=3$ in panel~(b1). Similarly, the observable $O$, whilst subject to fluctuations in both cases, is on average further from its limiting value of $1.0$ for $\mu=1$ in panel~(a2) compared to $\mu=3$ in panel~(b2). It is interesting to note that, despite the factorisation assumption being relatively inaccurate in this regime, the two-dimensional histograms are still well centred around the stable equilibrium. The two-dimensional histograms in the respective ranges of the $(\langle \hat{a}^{\dagger}_{1} \hat{a}_{1} \rangle,\langle \hat{a}^{\dagger}_{2} \hat{a}_{2} \rangle)$-plane in \fref{fig:Quant_Int_i}(a3) and (b3) show  distributions that are well centered around the stable equilibrium on the symmetry axis. Note that the colour map is scaled to account for the dependence of the size of the bins on $\mu$; this also allow for a direct comparison of the (relative) heights of the histograms for different values of $\mu$. There is considerable spread due to the fluctuations in the quantum trajectories, which are indeed comparable with the temporal traces in panels~(a1) and (b1). The spread is smaller and more symmetrical around the equilibrium for $\mu=3$ in panel~(b3) compared to that for $\mu=1$ in panel~(a3). This illustrates that the distribution becomes more Gaussian with a smaller variance as $\mu$ is increased.

\begin{figure}[t!]
   \centering
   \includegraphics{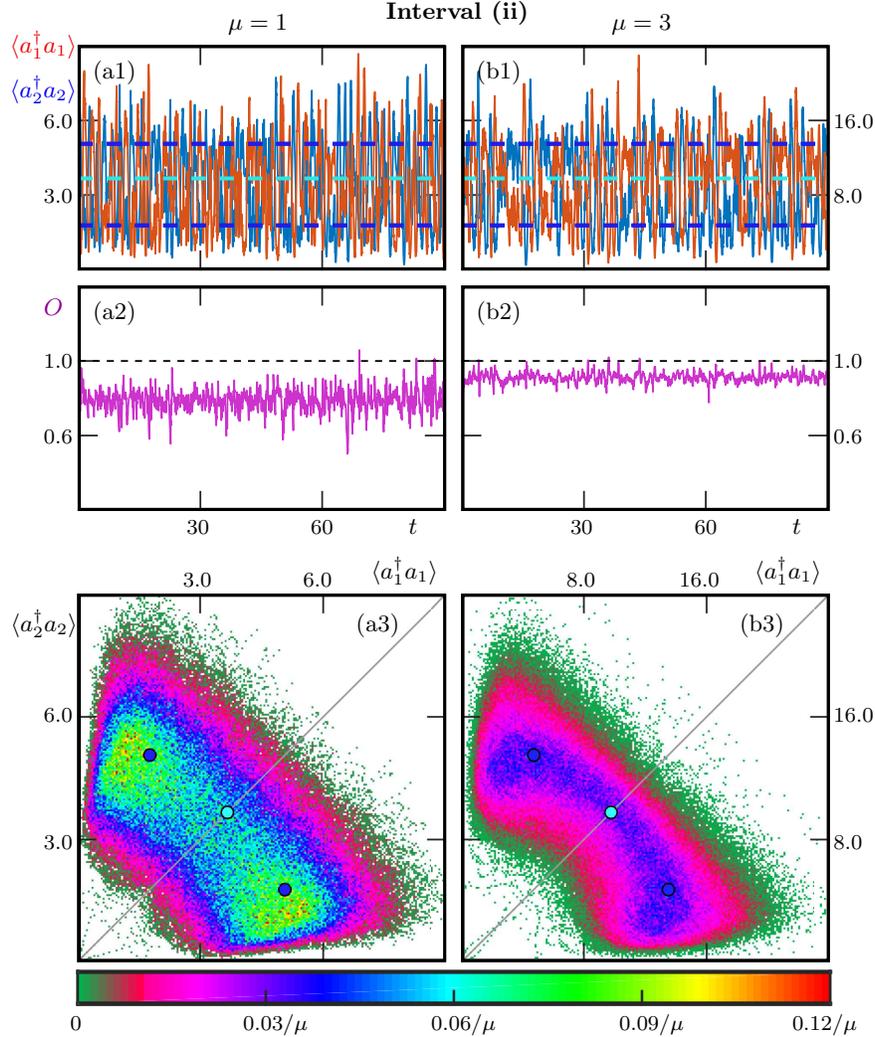}
    \caption{Quantum trajectories of system~\eqref{eq:LindForm} in interval~\textbf{(ii)} for $F_\mu/\sqrt{\mu} = f = 4.0$ as represented in \fref{fig:Quant_Int_i}.}
\label{fig:Quant_Int_ii}
\end{figure}

In interval~\textbf{(ii)}, past the pitchfork bifurcation $\mathbf{P_1}$, the symmetric equilibrium of system~\eqref{eq:CoupledAB} is now unstable, and there is a pair of stable asymmetric equilibria, each with their own basin of attraction. \Fref{fig:Quant_Int_ii} shows that quantum trajectories of~\eqref{eq:LindForm} display switching between the two asymmetric states in between (quite short) epochs of localisation near one of them. The switching appears to be dominant in the temporal trace for $\mu=1$ in panel~(a1), and epochs of localisation (while still short) are visible more clearly for $\mu=3$ in panel~(b1). Overall, the role of fluctuations seems to be much more important here than in interval~\textbf{(i)}, as they drive switching between the two stable solutions. These observations are represented in the two-dimensional histograms in panels~(a3) and (b3) by the fact that the distributions are now bimodal, quite broad and not sharply focused around the two stable asymmetric equilibria. Notice also the existence of a `bridge' between the areas of localisation near the asymmetric equilibria, which reflects the likely route for the switching driven by the quantum fluctuations. Again, the histogram is less broad, and its features are crisper for $\mu=3$ in panel~(b3) compared to $\mu=1$ in panel~(a3). 

\begin{figure}[t!]
   \centering
   \includegraphics{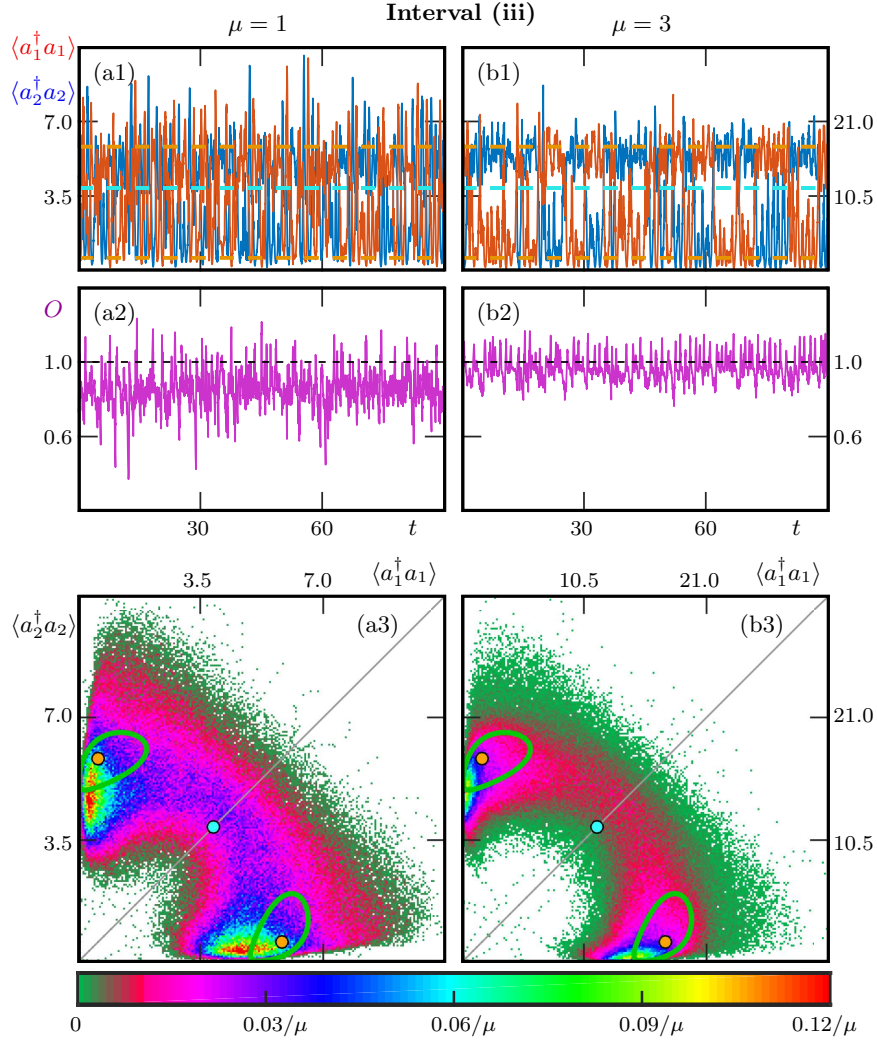}
    \caption{Quantum trajectories of system~\eqref{eq:LindForm} in interval~\textbf{(iii)} for $F_\mu/\sqrt{\mu} = f = 6.0$ as represented in \fref{fig:Quant_Int_i}; also shown in the histograms is the pair of stable asymmetric periodic solutions (green closed curves).}
\label{fig:Quant_Int_iii}
\end{figure}

\Fref{fig:Quant_Int_iii} illustrates the situation in interval~\textbf{(iii)}, in between the two Hopf bifurcation points $\mathbf{H_1}$ and $\mathbf{H_2}$, with the new feature of a pair of attracting periodic solutions of system~\eqref{eq:CoupledAB} near the now unstable asymmetric equilibria. The quantum trajectories of~\eqref{eq:LindForm} still display switching between these two asymmetric periodic states with epochs near one of them, with a considerable level of fluctuations. As before, the level of fluctuations is relatively higher for $\mu=1$ in panels~(a1) and~(a2) than for $\mu=3$ in panels~(b1) and~(b2). Comparison with \fref{fig:Quant_Int_ii} shows that the fluctuations are larger compared to the situation in interval~\textbf{(ii)}. In particular, the fluctuations during epochs of localisation are now larger since the quantum trajectories are no longer attracted to a steady state. The histograms in \fref{fig:Quant_Int_iii}(a3) and~(b3) show distributions that illustrate the properties of quantum trajectories differently. While the maxima are near the stable periodic solutions of the semiclassical ODE, there appears to be no definite fingerprint of periodicity of the quantum trajectories in the $(\langle \hat{a}^{\dagger}_{1} \hat{a}_{1} \rangle,\langle \hat{a}^{\dagger}_{2} \hat{a}_{2} \rangle)$-plane. Notice that there is again a clear `bridge' of preferred switching, which is considerably sharper for $\mu=1$ compared to $\mu=3$, but the histograms do not appear to identify the pair of periodic attractors. 

\begin{figure}[t!]
   \centering
   \includegraphics{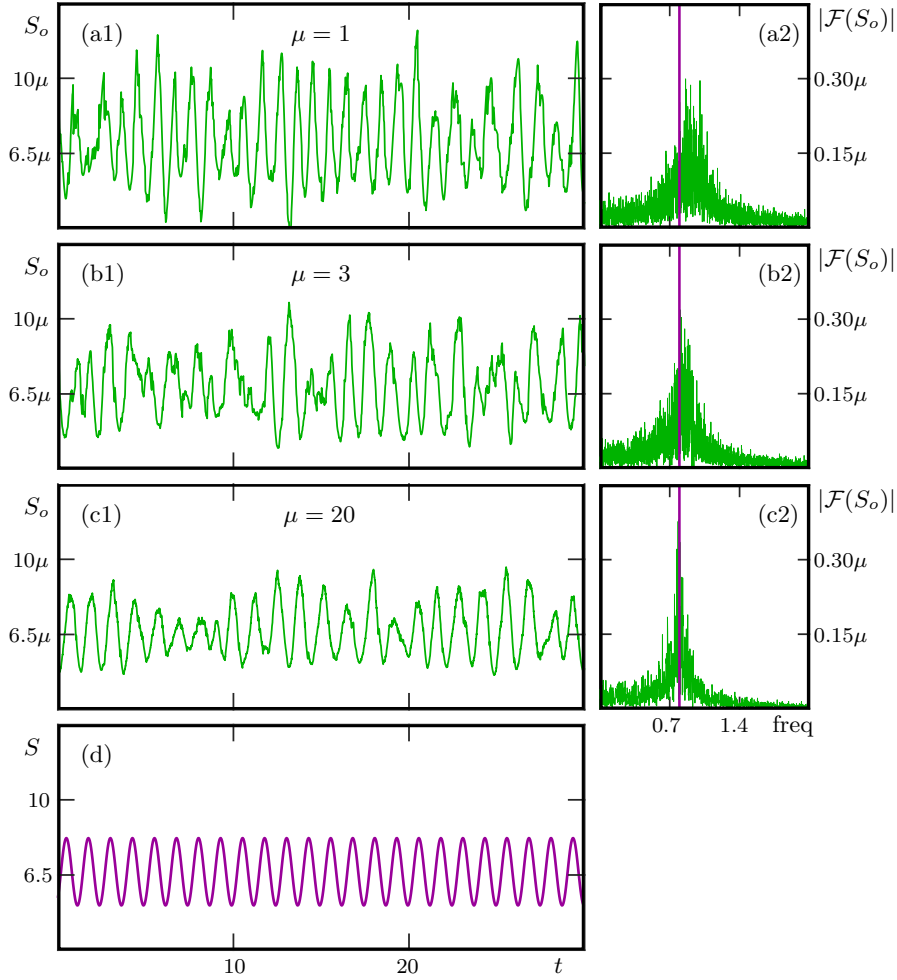}
    \caption{Comparison of oscillatory behaviour in interval~\textbf{(iii)} for $F_\mu/\sqrt{\mu} = f = 6.0$. Panels~(a1), (b1) and (c1) show temporal traces of the  observable $S_o=\langle \hat{a}^{\dagger}_{1} \hat{a}_{1} \rangle + \langle \hat{a}^{\dagger}_{2} \hat{a}_{2} \rangle$ of quantum trajectories of system~\eqref{eq:LindForm} with $\mu=1$, $\mu=3$ and $\mu=20$, respectively. Panels~(a2), (b2) and (c2) show the corresponding power spectra $|\mathcal{F}(S_o)|$ (green data); here the vertical purple line indicates the frequency of the corresponding periodic solution of system~\eqref{eq:LindForm}, which is shown in panel~(d) as a temporal traces of $S = |A|^2 + |B|^2$.}
\label{fig:Fourier}
\end{figure}

The above discussion shows that the frequent switching between the two localised oscillations obscures the possible periodicity of the observables $\langle \hat{a}^{\dagger}_{1} \hat{a}_{1} \rangle$ and $\langle \hat{a}^{\dagger}_{2} \hat{a}_{2} \rangle$. To identify the fingerprint of these semiclassical periodic oscillations in the quantum realm, we therefore now show in \fref{fig:Fourier} temporal traces of the observable $S_o=\langle \hat{a}^{\dagger}_{1} \hat{a}_{1} \rangle + \langle \hat{a}^{\dagger}_{2} \hat{a}_{2} \rangle$ of the quantum trajectories as well as their spectra, for $\mu=1$, $\mu=3$ and also for $\mu=20$. The observable $S_o$ (subject to the same scaling by $\mu$) is the quantum analogue of the total intensity $S = |A|^2 + |B|^2$ of the semiclassical ODE. Due to the symmetry properties of the Bose-Hubbard dimer, switching between symmetric states manifests itself much less in $S_o$. More specifically, due to its invariance under the permutation of the two sites, $S_o$ minimises fluctuation transients driven by the quantum system. This is why one can observe signs of periodicity in the temporal traces in \fref{fig:Fourier}(a1) to~(c1). The temporal trace of the semiclassical periodic orbit is shown in \fref{fig:Fourier}(d) for comparison; note that it is close to being sinusoidal, which is due to the periodic orbit still being close to the Hopf bifurcation. As $\mu$ is increased, fluctuations are reduced, and the periodicity in the temporal trace of $S_o$ becomes crisper. Indeed, the temporal trace for $\mu=20$ in panel~(c1) is recognised as a `noisy version' of the periodic signal in panel~(d) and, hence, clearly contains fingerprints of the semiclassical periodic solution. This observation is quantified by power spectra $|\mathcal{F}(S_o)|$ of the respective temporal traces. Already for $\mu=1$ in panel~(a2) the spectrum shows a recognisable peak near the main frequency of the semiclassical periodic temporal trace. For $\mu=3$ in panel~(b2), the spectrum is sharper and its frequency closer to that of the semiclassical oscillation, and this is even more the case for $\mu=20$ in panel~(c2). We conclude that \fref{fig:Fourier} clearly shows the emergence of periodicity in the quantum realm provided $\mu$, that is, the photon number, is taken to be sufficiently large (but still moderate). We remark that this phenomenon has also been observed recently in quantum trajectory simulations of the unbalanced Dicke model \cite{PhysRevResearch.Kevin2021}.

\subsection{Comparison in intervals \textbf{(iv)} to \textbf{(vi)}}
\label{sec:Int_vi_to_vi}

\begin{figure}[t!]
   \centering
   \includegraphics{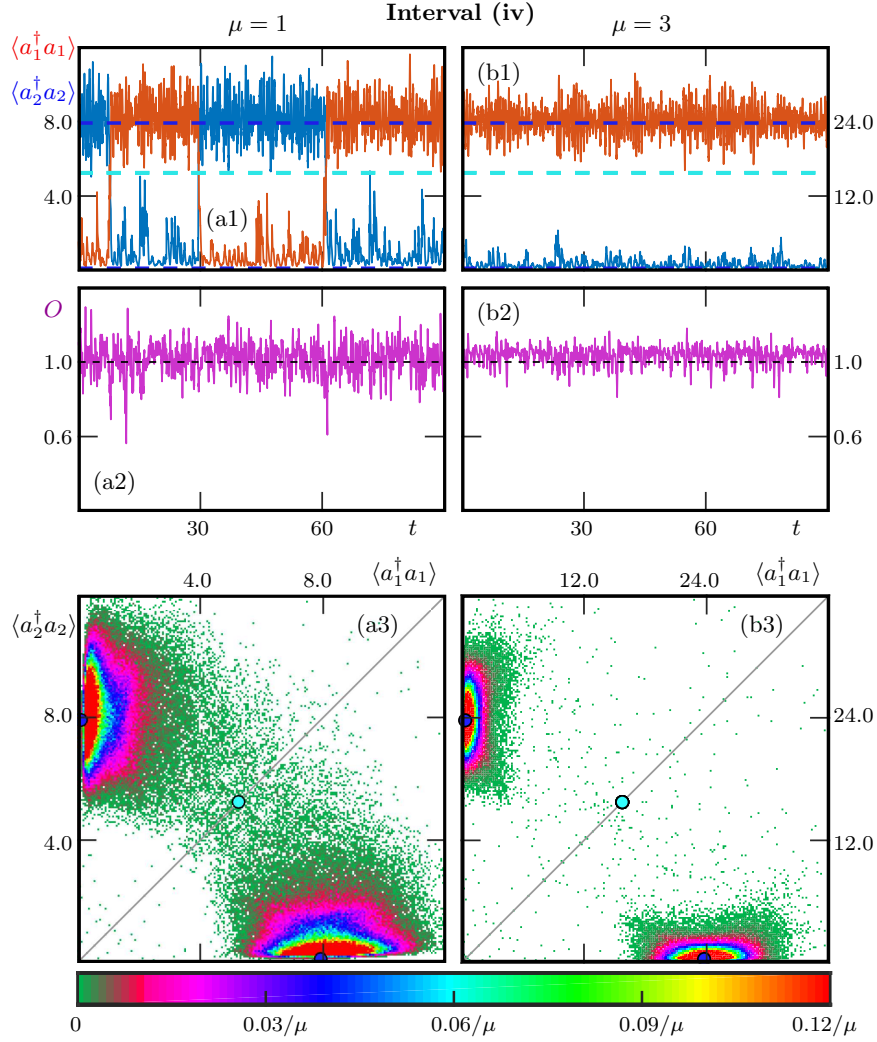}
    \caption{Quantum trajectories of system~\eqref{eq:LindForm} in interval~\textbf{(iv)} for $F_\mu/\sqrt{\mu} = f = 10.0$ as represented in \fref{fig:Quant_Int_i}.}
\label{fig:Quant_Int_iv}
\end{figure}

In interval~\textbf{(iv)}, past the second Hopf bifurcation $\mathbf{H_2}$, the two asymmetric equilibria are again stable and the only attractors of system~\eqref{eq:CoupledAB}. In other words, the situation is topologically the same as in interval~\textbf{(ii)}. However, as \fref{fig:Quant_Int_iv} shows, we find marked differences in the observed behaviour of quantum trajectories of~\eqref{eq:LindForm}. The temporal trace for $\mu=1$ in panel~(a1) show long epochs where the quantum trajectory is localised near one of the stable steady states, with much more occasional switchings between them. For $\mu=3$ in panel~(b1), the relative strength of quantum fluctuations is now so low that not a single switching occurs in the time window presented. Comparison with \fref{fig:Quant_Int_ii} shows that the overall level of fluctuation is comparable in intervals~\textbf{(ii)} and~\textbf{(iv)}. However, the observable $O$ is now seen to fluctuate around unity in panels~(a2) and~(b2), meaning that the factorization assumption is reasonable in interval~\textbf{(iv)}. Notice further that switching events manifest themselves in the observable $O$ in panel~(a2) as sudden larger spikes away from its average. The clear observation that quantum trajectories linger much longer near one of the two stable equilibria in interval~\textbf{(iv)} is explained by the fact that these equilibria are more attracting and also further apart from each other and from the unstable symmetric equilibrium from which they bifurcated. More specifically, the two stable equilibria represent a situation of extreme symmetry breaking where one site has practically all photons of the overall coupled system, while the other has near-zero photons --- this in spite of the fact that both sites are pumped symmetrically. This distance between the two attractors and the increased residence times of quantum trajectories are illustrated very clearly by the histograms in panels~(a3) and~(b3). Already for $\mu=1$ in panel~(a3), the distribution is very bimodal with pronounced peaks near the two attracting equilibria system~\eqref{eq:CoupledAB}, which are very close to the coordinate axes owing to the fact that one of the two intensities is practically zero; the `bridge' corresponding to preferred switching is now much weakened. For $\mu=3$ in panel~(b3), there is no longer a discernible bridge due to the occurrence of very few transitions between the two attractors. The histogram is now quite sharply focused on the pair of stable asymmetric equilibria, with very well defined peaks. That notwithstanding, there are still sufficiently many switchings due to quantum fluctuations to ensure that the computed histogram captures both attractors.

\begin{figure*}
   \centering
   \includegraphics{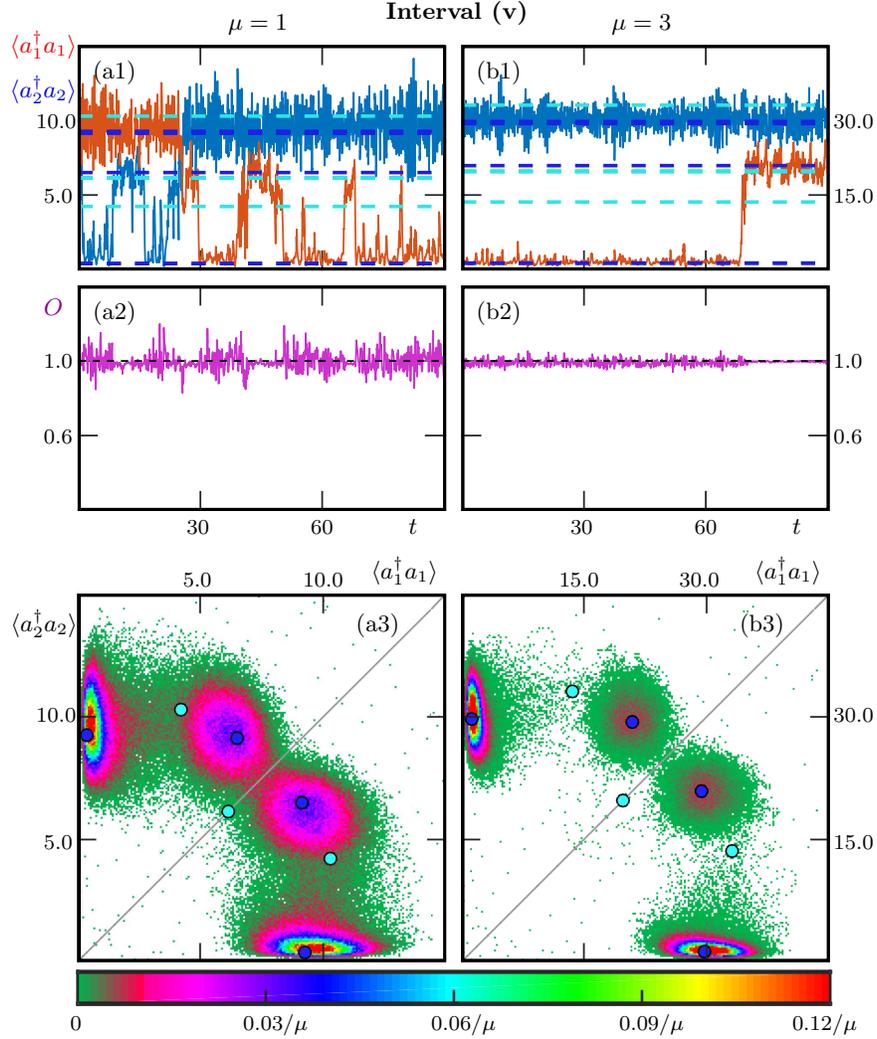}
    \caption{Quantum trajectories of system~\eqref{eq:LindForm} in interval~\textbf{(v)} for $F_\mu/\sqrt{\mu} = f = 13.0$ as represented in \fref{fig:Quant_Int_i}.}
\label{fig:Quant_Int_v}
\end{figure*}

The behaviour in interval~\textbf{(v)}, in between the two saddle-node bifurcations $\mathbf{S^*_1}$ and $\mathbf{S^*_2}$ is characterised by the existence of an additional pair of stable asymmetric equilibria of system~\eqref{eq:CoupledAB}, as well as a pair of unstable asymmetric equilibria. This means that there are now a total of four attractors with their respective basins of attraction.  The additional stable equilibria are characterised by a relatively small imbalance between the sites compared to the other pair, which still represent the scenario where practically all photons are at one of the two sites. \Fref{fig:Quant_Int_v} shows that quantum trajectories of~\eqref{eq:LindForm} remain localised near one of the four stable equilibria for a certain amount of time and then switch to being localised near another stable equilibrium and so on. The temporal trace for $\mu=1$ in panel~(a1) shows quite a number of switchings. While for $\mu=3$ in panel~(b1), residence times are longer, and the number of switchings per time interval is decreased. Interestingly, the overall level of fluctuation is relatively quite low compared to earlier cases. Note also that, as seen in panels~(a2) and ~(b2), localisation near the new equilibria with smaller values of the intensity is associated with especially low fluctuations and around an average value of $O=1.0$, implying the factorization assumption is accurate in this regime. These attractors feature significant photon numbers in both cavities, such that the role of quantum fluctuations is less important. However, whilst the other attractors feature a large population in one cavity, the other, almost empty, cavity is strongly impacted by fluctuations. The four different attractors are also clear features of the histograms in panels~(a3) and~(b3). The distributions show clear peaks near each of the stable equilibria, which are considerable crisper for $\mu=3$ compared to $\mu=1$. Moreover, the histograms are considerably larger near the pair of equilibria where one of the intensities is practically zero. Note that switching events follow `weak bridges' between the attractors, meaning that switching between neighbouring attractors in the $(\langle \hat{a}^{\dagger}_{1} \hat{a}_{1} \rangle,\langle \hat{a}^{\dagger}_{2} \hat{a}_{2} \rangle)$-plane are by far the ones that are most likely to occur.

\begin{figure}[t!]
   \centering
   \includegraphics{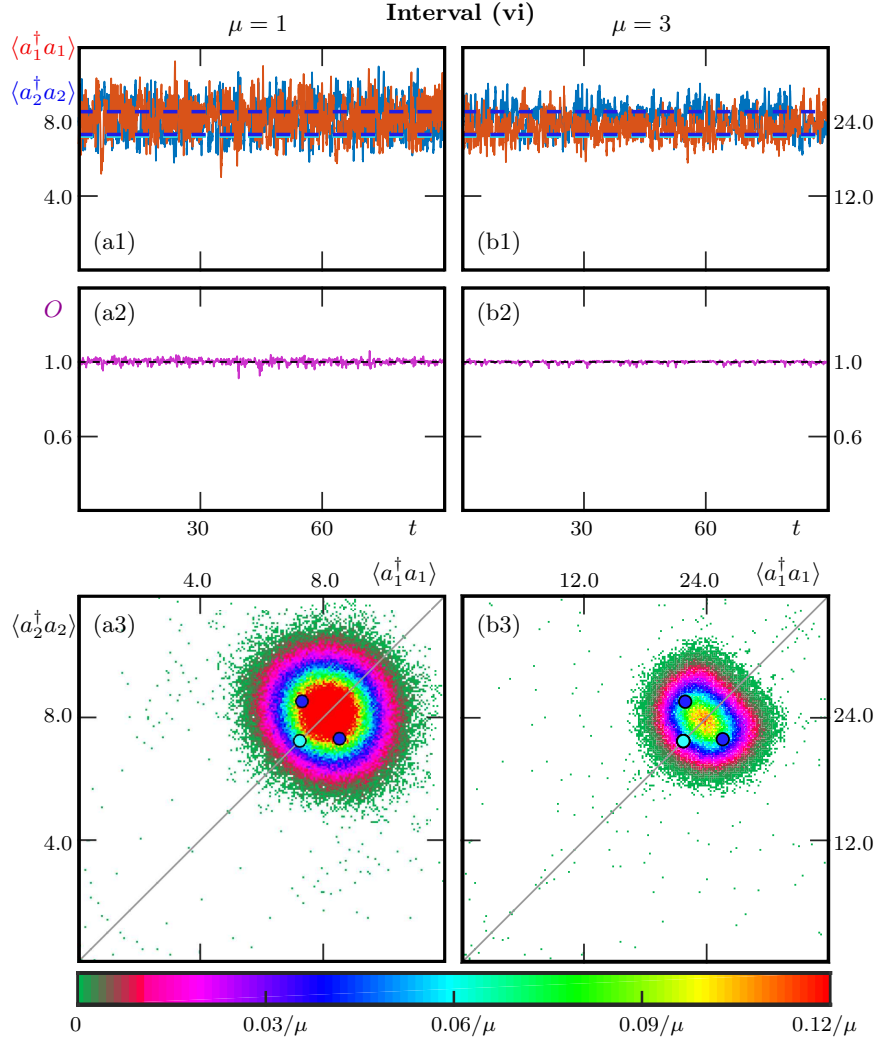}
    \caption{Quantum trajectories of system~\eqref{eq:LindForm} in interval~\textbf{(vi)} for $F_\mu/\sqrt{\mu} = f = 17.0$ as represented in \fref{fig:Quant_Int_i}.}
\label{fig:Quant_Int_vi}
\end{figure}

\Fref{fig:Quant_Int_vi} shows the situation in interval~\textbf{(vi)}, where there are again only two stable asymmetric equilibria of~\eqref{eq:CoupledAB}, namely the ones near the symmetry line with a relatively small imbalance between the sites. As panels~(a1) and~(b1) show, the quantum trajectories of~\eqref{eq:LindForm} remain near both of these equilibria, with negligible residence times near each attractor and many switchings per time interval. Residence times are notably larger for $\mu=3$ that for $\mu=1$, but remain very small for either case compared to the (topologically equivalent) situation in interval~\textbf{(iv)} in \fref{fig:Quant_Int_iv}. The level of fluctuations appears to be relatively small, and the observable $O$ in \fref{fig:Quant_Int_vi}(a2) and~(b2) remains very close to $1.0$ throughout; again, any noticeable fluctuations of $O$ appear to be due to switchings between the two stable equilibria. However, the role of these fluctuations is important, as the frequent switching means that the associated histograms in panels~(a3) and~(b3) are characterised by a single peak with a maximum at a point in between the two stable asymmetric equilibria. In other words, the two nearby attractors are not resolved: while the histogram for $\mu=3$ in panel~(b3) appears to be more elongated around the two attracting equilibria, it still has only a single peak. This is a direct reflection of the very low residence times of localisation of as measured by the observables $\langle \hat{a}^{\dagger}_{1} \hat{a}_{1} \rangle$ and $\langle \hat{a}^{\dagger}_{2} \hat{a}_{2} \rangle$. As we will see below, it is possible to detect the (weak) localisation of the quantum trajectories already for $\mu=3$, namely by considering the difference between these two observables; see already \fref{fig:Violin}(b2).

\section{Evolution of histograms with  $F_{\mu}$}
\label{sec:violin}

\begin{figure}[t!]
   \centering
   \includegraphics{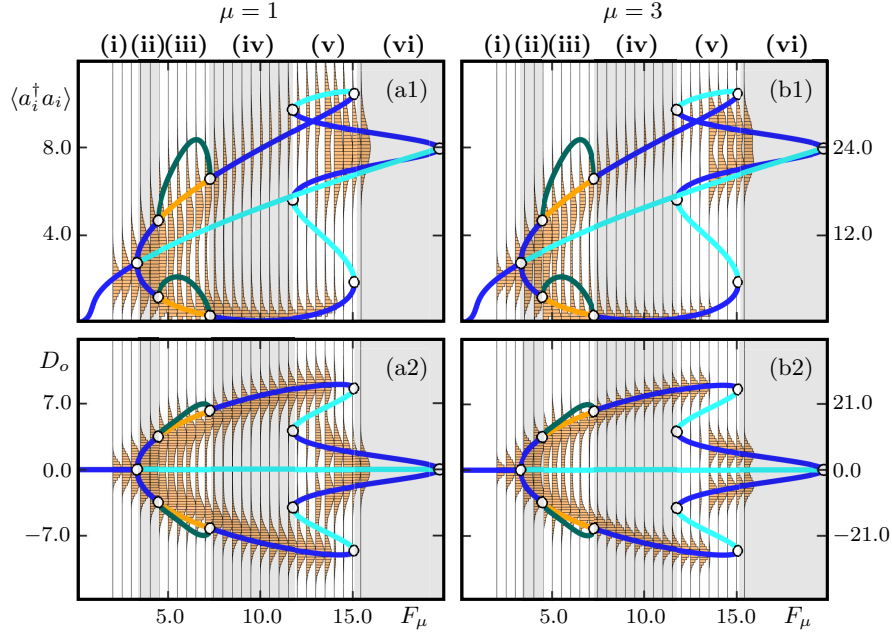}
    \caption{Scaled histograms of system~\eqref{eq:LindForm} for $\mu=1$ (left column) and for $\mu=3$ (right column) of the observables $\langle \hat{a}^{\dagger}_{i} \hat{a}_{i} \rangle$ (top row) and of their difference $D_o=\langle \hat{a}^{\dagger}_{1} \hat{a}_{1} \rangle - \langle \hat{a}^{\dagger}_{2} \hat{a}_{2} \rangle$ (bottom row) at different values of $F_\mu$ superimposed on the bifurcation diagram of system~\eqref{eq:CoupledAB}. Here $F_{\mu}$ is increased from $2.0\mu$ to $15.75\mu$ in steps of $0.5\mu$, and the histograms are scaled to have a maximum of $0.5\mu$. The contrast in the background shading of the panels highlights intervals~\textbf{(i)} to~~\textbf{(vi)}.}
\label{fig:Violin}
\end{figure} 

\Fref{fig:Violin} provides an overview of how the statistical properties of quantum trajectories of system~\eqref{eq:LindForm} develop with the pump strength $F_\mu$, and how this compares with the corresponding bifurcation diagram of the semiclassical ODE~\eqref{eq:CoupledAB}. Here, we again consider the two cases $\mu=1$ and $\mu=3$. In this representation, histograms are shown for $F_{\mu}$ starting from $2.0\mu$ in steps of $0.5\mu$ and up to $15.75\mu$, where the maximum of each histogram is scaled to the $F_{\mu}$-size of $0.5\mu$. As for the two-dimensional histograms shown in previous figures, these histograms are constructed from three different realisations of quantum trajectories by sampling 50,000 equidistant points in the time interval $(0,10^4)$, where we now use 80 uniform bins over the shown range of the respective observable. The invariance of system~\eqref{eq:LindForm} under the permutation of the sites is again used to double the number of points and symmetrise the data; the maxima are then scaled to $0.5\mu$.

\Fref{fig:Violin} shows histogram plots for two different observables. Panels~(a2) and~(b2) show histograms for $\langle \hat{a}^{\dagger}_{i} \hat{a}_{i} \rangle$, which corresponds to the projection of the respective two-dimensional histogram onto the $\langle \hat{a}^{\dagger}_{1} \hat{a}_{1} \rangle$-axis (or equivalently the $\langle \hat{a}^{\dagger}_{2} \hat{a}_{2} \rangle$-axis). To obtain additional information regarding semiclassical fingerprints in the distributions, \fref{fig:Violin}(a2) and~(b2) show histograms for the difference $D_o=\langle \hat{a}^{\dagger}_{1} \hat{a}_{1} \rangle - \langle \hat{a}^{\dagger}_{2} \hat{a}_{2} \rangle$; note that considering this observable corresponds to the projection of the respective two-dimensional histogram onto the antidiagonal and implies the symmetry of panels~(a2) and~(b2) with respect to the $F_{\mu}$-axis.

The evolution of the histograms for $\mu=1$ is shown in \fref{fig:Violin}(a1) and~(a2). There are noticeable changes in the statistical properties of the two observables associated with the transitions through the different bifurcations, while the distributions remain more or less the same in the intervals~\textbf{(i)} to~\textbf{(vi)} that are covered by the shown range of $F_{\mu}$; compare with \fref{fig:BifDiag}. Starting at low $F_{\mu}$ in interval~\textbf{(i)}, one first observes that the histogram widens as the first pitchfork bifurcation is approached and subsequently becomes bimodal in interval~\textbf{(ii)}, with a pair of peaks near the two stable asymmetric equilibria. Notice the large component of the distribution in between these two equilibria, which corresponds to frequent switchings between them. This `bridge' and, hence, the number of switchings in the time interval clearly become much less pronounced past interval~\textbf{(iii)}. The existence of stable periodic orbits, on the other hand, is not evident in the histograms. When the second Hopf bifurcation is reached, the distribution is strongly bimodal and remains clearly localised near the re-stabilised asymmetric equilibria throughout interval~\textbf{(iv)}. Notice that in \fref{fig:Violin}(a1) for the observable $\langle \hat{a}^{\dagger}_{i} \hat{a}_{i} \rangle$ the solution is less well resolved near the upper branch compared to the lower branch, due to the larger fluctuations for the site with more photons; this issue clearly does not arise for the symmetric observable $D_o$ in panel~(a2). The strong localisation extends well into interval~\textbf{(v)} with the additional pair of stable equilibria. There is a somewhat gradual change of the histogram to localisation around the other pair of stable equilibria in this interval. Notice that the distinction between these two equilibria is quite weak initially and quickly becomes nonexistent, with a distribution with a maximum in between the two attractors, even for the observable $D_o$.

The evolution of the histograms for $\mu=3$ in \fref{fig:Violin}(b1) and~(b2) is quite similar, but the distributions are more clearly resolved, that is, more concentrated at the respective attractor. However, there are noticeable differences in the upper range of $F_{\mu}$. In interval~\textbf{(v)}, the switching to a different pair of attractors now manifests itself as a quite sudden change of the histogram; this reflects the scarcity of switchings in a finite time series. Moreover, the two new asymmetric equilibria are now distinguished by the histogram, especially clearly for the observable $D_o$ in panel~(b2). Overall, we conclude that the histogram plots of \fref{fig:Violin}, especially those in panels~(a2) and~(b2) for the symmetric observable $D_o$, provide a good summary of how the statistical properties of quantum trajectories change both with the pump strength $F_\mu$ as well as with increasing numbers of photons as represented by the scaling parameter $\mu$. Indeed, this representation agrees with the results in \sref{sec:ComF} regarding the behaviour for representative values of $F_\mu$ in the intervals~\textbf{(i)} to~\textbf{(vi)} --- but it also provides insight into how the distributions of quantum trajectories change from interval to interval, as semiclassical bifurcations are encountered.

\section{Antibunching and entanglement}

\begin{figure}[t!]
   \centering
   \includegraphics{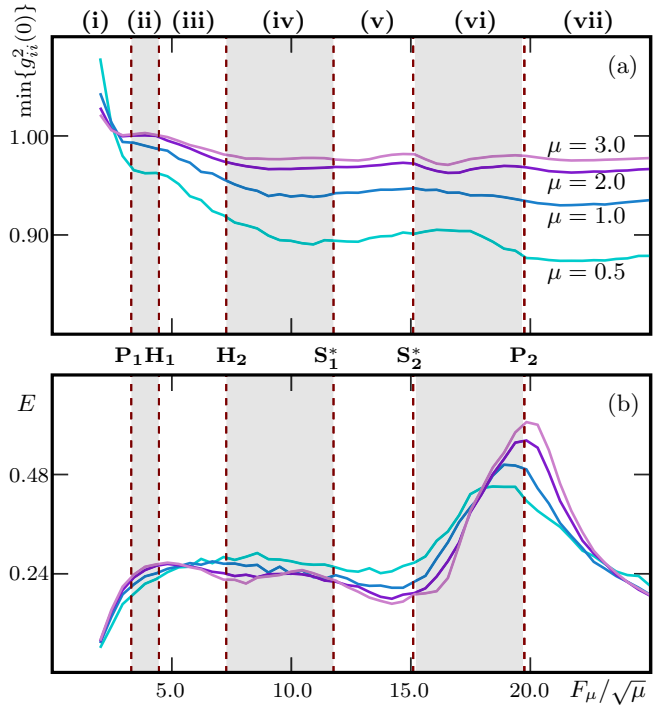}
    \caption{The minimum $\min\{g_{11}^{(2)}(0), g_{22}^{(2)}(0)\}$ of the averaged $g^{(2)}$-functions of the two sites~(a) and the averaged von Neumann entropy $E$~(b) as a function of pump strength $F_\mu$ as functions of the pump strength $F_\mu$. Each curve was computed from a single ramped quantum trajectory of system~\eqref{eq:LindForm}, where $\mu =0.5$ (cyan), $\mu =1.0$ (blue), $\mu =2.0$ (purple), and $\mu =3.0$ (lilac). Red-dashed vertical lines are at the bifurcations of~\eqref{eq:CoupledAB}, and the corresponding intervals~\textbf{(i)} to~\textbf{(vii)} are highlighted by contrasting background shading.}
\label{fig:AntiBunEnt}
\end{figure} 

We now investigate whether there is evidence of quantum phenomena in system~\eqref{eq:LindForm} for low photon numbers as the pump strength $F_\mu$ is varied. To this end, \fref{fig:AntiBunEnt} shows the two indicator functions $\min\{g_{11}^{(2)}(0), g_{22}^{(2)}(0)\}$ and $E$ over the relevant range of $F_\mu$, where the semiclassical bifurcations are shown and the associated intervals~\textbf{(i)} to~\textbf{(vii)} of different behaviour are highlighted.

Antibunching of photons refers to the emission of a photon reducing the probability of a subsequent emission, and is a strictly quantum phenomena. In the present context, it can be detected by the condition that $\min\{g_{11}^{(2)}(0), g_{22}^{(2)}(0)\}<1$, where the function $g^{(2)}_{ii}(0)$ is the second-order correlation function for the light in cavity $i$
$$g^{(2)}_{ii}(0) = \frac{\langle \hat{a}^\dagger_i \hat{a}^\dagger_i \hat{a}_i \hat{a}_i \rangle}{\left(\langle \hat{a}^\dagger_i \hat{a} \rangle\right)^2}.$$ 
Plotting the smaller of the time averaged second-order correlation functions of the two sites, $g^{(2)}_{11}(0)$ and $g^{(2)}_{22}(0)$, for a single quantum trajectory allows us to identify antibunching \cite{PhysRevLett.108.183601}. \Fref{fig:AntiBunEnt}(a) shows time-averaged values of $\min\{g_{11}^{(2)}(0), g_{22}^{(2)}(0)\}$ for $\mu =0.5$, $\mu =1.0$, $\mu =2.0$, and $\mu =3.0$ at different values of $F_\mu$ for single trajectories of length $t = 500$. In interval~\textbf{(i)}, both sites display slight bunching, less so for larger $\mu$, implying a thermal photon number distribution. Past the pitchfork bifurcation~$\mathbf{P_1}$ and the appearance of asymmetric attracting states of system~\eqref{eq:CoupledAB}, we find $\min\{g_{11}^{(2)}(0), g_{22}^{(2)}(0)\}<1$: consistently in intervals~\textbf{(iii)} to~\textbf{(vii)} for all shown values of $\mu$, and including interval~\textbf{(ii)} for $\mu =0.5$ and $\mu =1.0$. Hence, the site with more photons is always anti-bunched from about $F_\mu = 5.0$ and higher. As one would expect, antibunching is reduced as $\mu$ increases and the closer the system is to its semiclassical limit, and $\min\{g_{11}^{(2)}(0), g_{22}^{(2)}(0)\}$ appears to approach the value $1.0$ as $\mu$ is increased. 

To investigate the existence of entanglement between the sites, we consider the time-averaged von Neumann entropy of one of the cavities, given by~\cite{PhysRevLett.2004_Howard,PhysRevLett.88.197901}
$$E = - \overline{\mbox{trace}(\rho_1(t) \ln \rho_1(t))},$$
where $\rho_1(t)$ is the density matrix for site 1 at time $t$ (or equivalently site 2) and the overline represents time averaging. \Fref{fig:AntiBunEnt}~(b) shows $E$ as a function of $F_\mu$ for $\mu =0.5$, $\mu =1.0$, $\mu =2.0$, and $\mu =3.0$, as computed from the same quantum trajectories described above. Entanglement is identified by the condition that $E>0$, which means that the states are not `pure', and we conclude that there is entanglement throughout the entire range of $F_\mu$. Notice that the level of entanglement as measured by $E$ increases in interval~\textbf{(i)} and then, past the first pitchfork bifurcation~$\mathbf{P_1}$, reaches and stays on a plateau throughout interval~\textbf{(ii)} to~\textbf{(v)}. The indicator $E$ then increases quite steeply in interval~\textbf{(vi)} with maximal multistability between asymmetric attractors, reaches a maximum near the second pitchfork bifurcation~$\mathbf{P_2}$ and then decreases equally steeply in interval~\textbf{(vii)} where the symmetric equilibrium is again the only attractor of the limiting system~\eqref{eq:CoupledAB}. The von Neumann entropy in the plateau appears to show slowly reducing entanglement as $\mu$ increases, suggesting that entanglement goes away in the thermodynamic limit. However, the maximum of $E$ appears to become larger and more narrow with increasing $\mu$. It is a known phenomenon that, near certain types of phase transitions, there is a divergence in entanglement at the critical point in the thermodynamic limit~\cite{Osterloh:2002wc,PhysRevLett.92.073602,Reslen_2005,PhysRevA.71.064101}. Whether this is the explanation for the sharpening of the von Neumann energy near $\mathbf{P_2}$ remains an interesting question for future research.

\section{Conclusions and outlook}
\label{sec:concl}

The case study of the open two-site Bose-Hubbard dimer presented here shows that it is possible to identify recognisable fingerprints of phase transitions --- that is, bifurcations --- of the limiting semiclassical (mean-field) model in the quantum realm. More specifically, we considered the case of negative intermode coupling when the semiclassical model features a transition, as the pump strength is increased, from symmetric dynamics via symmetry breaking at a pitchfork bifurcation to oscillatory dynamics and then to multistability between different types of asymmetric states. These features were recognised reliably in the statistical properties of different observables of quantum trajectories, even quite far from the semiclassical limit, that is, for low numbers of photons at each site. 

These theoretical results are fundamental in nature and suggest that finding quantum signatures of such rather complex nonlinear phenomena may be possible in an experimental context. For the system under study here, this has been achieved up to a moderate level of the pump strength, allowing for the experimental verification of spontaneous symmetry breaking in good agreement with a bifurcation study of the semiclassical model \cite{GBKYA_2020}. We believe that the experimental identification of more complicated dynamics with additional levels of multistability --- including that involving different types of localised and non-localised chaotic dynamics  \cite{GBK_2021} --- should be possible. For the Bose-Hubbard dimer or other quantum systems such as the Dicke model, such experiments are challenging but would present an opportunity for studying how semiclassical chaos arises in quantum systems~\cite{PhysRevA.34.482,PhysRevLett.107.100401}.

\section*{Acknowledgement}
We thank Bruno Garbin, Ariel Levenson and Alejandro Yacomotti for sharing their insights on the open Bose--Hubbard dimer model, especially concerning their experiments. We also thank Ricardo Gutierrez-Jauregui for helpful comments.

\bibliographystyle{siam}
\bibliography{GMBK_QuanTraj_arXiv}

\end{document}